\documentclass[pra,twocolumn,showpacs]{revtex4}
\usepackage{amsmath}
\usepackage{amsfonts}
\usepackage{amssymb}
\usepackage{array}
\usepackage{floatflt}
\usepackage{graphicx}
\DeclareMathOperator{\tr}{Tr}
\newcommand*{\eq}[1]{Eq.~(\ref{#1})}
\newcommand*{\eqs}[2]{Eqs.~(\ref{#1})--(\ref{#2})}

\renewcommand{\l}{\left(}
\renewcommand{\r}{\right)}

\begin{document}
\title{Entanglement in alternating open spin--1/2 chains with XY-Hamiltonian}
\author{S.I.Doronin, E.B.Fel'dman, A.N.Pyrkov}
\affiliation{Institute of Problems of Chemical Physics,
Chernogolovka, Moscow Region, Russia, 142432}

\date{\today}

\begin{abstract}
We investigate entanglement of spin pairs in  alternating open spin
chains in the equilibrium state in an external magnetic field. We
calculate the reduced density matrix of spin pairs and estimate the
concurrence with Wootter's criteria. The obtained results
demonstrate the dependence of the entanglement on the temperature,
chain's length, the positions of the spins, and the ratio of the
spin-spin interaction constants.
\end{abstract}

\maketitle

\section{introduction}

A study of entangled states is an important direction in quantum
information theory. A quantitative investigation of the degree of
entanglement often poses considerable difficulties and is possible
only in relatively simple cases. The problem becomes much simpler,
if it is known how to diagonalize the
Hamiltonian~\cite{lieb,cruz,fbe}. In particular, many interesting
results were obtained for spin rings and infinite spin
chains~\cite{cw, amico, osborne}. On the other hand, possible
experimental realizations of qubits~\cite{levy} involve finite open
chains which were also used for the transfer of quantum states  from
one end of the chain to the other~\cite{kf, ekert}. Important
results were obtained by Wang~\cite{wang} who investigated open
boundary effects on the ground-state entanglement in a homogeneous
Heisenberg model. However, the models~\cite{cw, amico, osborne,
wang} do not allow one to solve the qubit addressing problem which
can be attacked in systems with several different Larmor
frequencies. Such Larmor frequency variations naturally emerge in
inhomogeneous systems with different constants of the spin-spin
interaction (SSI) for different pairs of neighboring spins.

In this paper, we investigate entanglement in open alternating spin
chains. Open alternating spin chains with XY--Hamiltonians can be
used as a first step for the solution of the qubit addressing
problem. Recently, the XY--Hamiltonian of an open finite alternating
spin chain has been diagonalized~\cite{fr,kf} on the basis of a
generalization of the classical methods~\cite{lieb} for homogeneous
chains. It was shown that the transfer of quantum states from one
end to the other is possible with $100\%$ fidelity for longer
alternating chains than in the case of homogeneous chains~\cite{kf}.

The concurrence~\cite{woot} is used as a quantitative measure of
entanglement. A possibility of the exact diagonalization of the
Hamiltonian~\cite{kf} makes it possible to construct the reduced
density matrix for any spin pair in a chain described by the
equilibrium density matrix. Then entanglement can be investigated
with Wootter's criteria~\cite{woot}. As a result, one can study the
influence of the temperature, the chain length, the distance from
the ends, and the ratio of the SSI constants on the entanglement of
any spin pair in the chain.

The present paper is organized as follows. In Section~{II} the model
Hamiltonian of the open alternating spin chain is described and the
eigenvalues of the Hamiltonian and its eigenvectors are given. In
Section~{III} we obtain the reduced density matrix for all pairs of
the neighboring spins of the chain. We study the dependence of the
spin pair entanglement on the temperature, chain's length, spin
positions and ratio of the SSI constants in Section~{IV}.

\section{the model}

We consider an open alternating spin--1/2 chain with the
XY--Hamiltonian in a strong external magnetic field. The Hamiltonian
of the system can be written as
\begin{multline}
\label{iham}
H=\sum_{n=1}^N\omega_nI_{nz}\\
+\sum_{n=1}^{N-1}D_{n,n+1}(I_{n,x}I_{n+1,x}+I_{n,y}I_{n+1,y}).
\end{multline}
Here,  $I_{n\alpha}$ are the spin--1/2 matrices $(\alpha=x,y,z),$
and $N$ is the number of the spins. The alternating Larmor
frequencies are determined by the equation
\begin{equation}
\label{altw}
\omega_n=\begin{cases} \omega_1,\qquad n\;\; is\;\; odd\\
\omega_2,\qquad n\;\; is\;\; even.
\end{cases}
\end{equation}
The alternating coupling constants are
\begin{equation}
\label{altd}
D_{n,n+1}=\begin{cases} D_1,\qquad n\;\; is\;\; odd\\
D_2,\qquad n\;\; is\;\; even.
\end{cases}
\end{equation}
We assume that the number of the spins, $N,$ is odd. The Hamiltonian
of \eq{iham} can be diagonalized~\cite{fr,kf} on the basis of a
generalization of the classical methods~\cite{lieb} for homogeneous
chains.

The Jordan--Wigner transformation~\cite{lieb,cruz} that  maps spins
onto free spinless fermions can be written as
\begin{gather}
I_{n,-}=I_{n,x}-iI_{n,y}=(-2)^{n-1}\l\prod_{l=1}^{n-1}I_{l,z}\r c_n,\\
I_{n,{+}}=I_{n,x}+iI_{n,y}=(-2)^{n-1}\l\prod_{l=1}^{n-1}I_{l,z}\r c_n^{\dag},\\
I_{n,z}=c_n^{\dag}c_n-1/2,
\end{gather}
where $c_n^{\dag}$ and $c_n$ are creation and annihilation
operators. As a result, the Hamiltonian in the matrix notation has
the following form
\begin{equation}
\label{mham} H=\frac12{\bf c^{\dag}}(D+2\Omega){\bf
c}-\frac12\sum_{n=1}^N\omega_n.
\end{equation}
Here we denote the row vector ${\bf
c^\dag}=(c_1^\dag,\ldots,c_N^\dag),$ the column vector ${\bf
c}=(c_1,\ldots,c_N)^t$ (the superscript $t$ represents the
transposition) and $D+2\Omega$ is a three-diagonal matrix.

The diagonalization of the matrix $D+2\Omega$ is performed by the
unitary transformation
\begin{equation}
D+2\Omega=U\Lambda~U^\dag,\quad
\Lambda={\textrm{diag}}\{\lambda_1,\ldots,\lambda_N\}.
\end{equation}
The new fermion operators $\gamma_k^\dag$ and $\gamma_k,$ which are
introduced by the relationships
\begin{equation}
c_n^\dag=\sum_{k=1}^Nu_{n,k}^*\gamma_k^\dag,\quad
c_n=\sum_{k=1}^Nu_{n,k}\gamma_k,
\end{equation}
transform the Hamiltonian of Eq.~(\ref{mham}) into the Hamiltonian
\begin{equation}
\label{fham}
H=\frac12\sum_{k=1}^N\lambda_k\gamma_k^\dag\gamma_k-\frac12\sum_{n=1}^N\omega_n.
\end{equation}
This is a free fermion Hamiltonian with the energy levels
$1/2\lambda_k.$ Finally, the eigenvalues $\lambda_k$ and
eigenvectors $|u_k\rangle=(u_{1k},u_{2k},\ldots,u_{Nk})^t$ of the
Hamiltonian of Eq.~(\ref{fham}) can be written as~\cite{fr}
\begin{equation}
\label{lamb}
\lambda_k=\begin{cases}
\omega_1+\omega_2+\sqrt{(\omega_1-\omega_2)^2+D_1^2\Delta_k},\\
k=1, 2, \ldots,\frac{N-1}{2}\\
2\omega_1, \quad k=\frac{N+1}{2}\\
\omega_1+\omega_2-\sqrt{(\omega_1-\omega_2)^2+D_1^2\Delta_k},\\
k=\frac{N+3}{2}, \frac{N+5}{2}, \ldots,N
\end{cases}
\end{equation}
where
\begin{equation}
\label{delt}
\Delta_k=1+2\delta\cos{\left(\frac{2\pi{k}}{N+1}\right)}+\delta^2,\quad
\delta=D_2/D_1.
\end{equation}
For all the indices $k=1,\ldots,N$ except the index $k=(N+1)/2$ the
eigenvector $|u_k\rangle$ has the elements
\begin{equation}
\label{evec}
u_{j,k}=\begin{cases}
A_k\frac{D_1}{\lambda_k-2\omega_1}\left[\delta\sin\left(\frac{\pi k(j-1)}{N+1}\right)\right.\\
\left.+\sin\left(\frac{\pi k(j+1)}{N+1}\right)\right],\\
j=1,3,5,\ldots,N\\
A_k\sin\left(\frac{\pi kj}{N+1}\right),\\
j=2,4,\ldots,N-1
\end{cases}
\end{equation}
with the normalization coefficient
\begin{equation}
\label{Ak}
A_k=\frac{2|\lambda_k-2\omega_1|}{\sqrt{N+1}}\frac1{\sqrt{(\lambda_k-2\omega_1)^2+D_1^2\Delta_k}}.
\end{equation}
The elements of the eigenvector $|u_{(N+1)/2}\rangle$ read
\begin{equation}
u_{j,(N+1)/2}=\begin{cases} B(-\delta)^{(N-j)/2}, \quad &j=1,3,5,\ldots,N\\
0, &j=2,4,\ldots,N-1
\end{cases}
\end{equation}
with the normalization coefficient
\begin{equation}
\label{B}
B=\left(\frac{\delta^2-1}{\delta^{N+1}-1}\right)^{1/2}.
\end{equation}

For $D_1=D_2=D$ and $\omega_1=\omega_2=\omega_0$ we obtain a
homogeneous chain and the Hamiltonian of Eq.~(\ref{fham}) can be
written as~\cite{fbe}
\begin{equation}
H=\sum_{k=1}^N\lambda_k\gamma_k^\dag\gamma_k-\frac12N\omega_0,
\end{equation}
with $\lambda_k=D\cos{\frac{\pi k}{N+1}}+\omega_0$ and eigenvector
$|u_k\rangle$ has the elements
\begin{equation}
u_{j,k}=\left(\frac{2}{N+1}\right)^{1/2}\sin{\left(\frac{\pi
kj}{N+1}\right)},\quad j=1,2,\ldots,N.
\end{equation}
For homogeneous chains these expressions are valid both for odd and
even N.

\section{the reduced density matrix}

In this section we consider a one--dimensional many--spin system in
a thermodynamic equilibrium state. We describe an algorithm for
obtaining the reduced density matrix for any spin pair. This
algorithm will be applied in order to obtain the reduced density
matrix of the nearest--neighbor spins $i$ and $i+1$ for the open
alternating chain with an odd number of the spins.

The density matrix, $\rho$, in the thermodynamic equilibrium system
is
\begin{equation}
\label{imatr} \rho=\frac{e^{-\beta H}}{Z},
\end{equation}
where $\beta=\hbar/kT$, $T$ is the temperature and $Z=\tr\{e^{-\beta
H}\}$ is the partition function. The density matrix, $\rho$, can be
written as
\begin{equation}
\rho=\sum_{\xi_1,\xi_2,\ldots,\xi_N=0}^3\alpha_{12\ldots
N}^{\xi_1\xi_2\ldots\xi_N}x_1^{\xi_1}\otimes\ldots\otimes
x_N^{\xi_N},
\end{equation}
where $N$ is a number of spins in the system, $\xi_k
\quad(k=1,2,\ldots,N)$ is one of the values $\{0,1,2,3\},$
$x_k^0=I_k$ is the unit matrix of the dimension $2\times 2,$
$x_k^1=I_{kx},$ $x_k^2=I_{ky},$  $x_k^3=I_{kz},$ and
$\alpha_{12\ldots N}^{\xi_1\xi_2\ldots\xi_N}$ is a numerical
coefficient.

In order to obtain the reduced density matrix for spins $i, j,$ we
consider the system of all other spins as the environment. Averaging
the density matrix, $\rho,$ over the environment and taking into
account that
$\tr\{x_k^{\xi_k}\}=0\quad(k=1,2,\ldots,N;\,\xi_k=1,2,3)$ we find
for the reduced density matrix, $\rho,$ of spins $i$ and $j$ the
following expression
\begin{equation}
\label{rmatr}
\rho_{ij}=\sum_{\xi_i,\xi_j=0}^3\alpha_{ij}^{\xi_i\xi_j}x_i^{\xi_i}\otimes
x_j^{\xi_j},
\end{equation}
where the coefficient $\alpha_{ij}^{\xi_i\xi_j}$ is defined as
\begin{equation}
\label{alpha} \alpha_{ij}^{\xi_i\xi_j}=\frac{2^{N-2}\tr\{\rho
x_i^{\xi_i}x_j^{\xi_j}\}}{\tr\{(x_i^{\xi_i})^2(x_j^{\xi_j})^2\}}.
\end{equation}
The calculations of the traces in \eq{alpha} are performed in an
$2^N$--dimension space. Here, the density matrix, $\rho,$ and the
density matrix, $\rho_{ij},$ are normalized to unity.

Since $[H, I_z]=0$ one can see that
\begin{multline}
\alpha_{ij}^{01}=\alpha_{ij}^{10}=\alpha_{ij}^{02}=\alpha_{ij}^{20}=0,\\
\alpha_{ij}^{13}=\alpha_{ij}^{23}=\alpha_{ij}^{31}=\alpha_{ij}^{32}=0.
\end{multline}
Furthermore, $\alpha_{ij}^{12}=4\tr{\{\rho I_{ix}I_{jy}\}}=0,$
$\alpha_{ij}^{21}=4\tr{\{\rho I_{iy}I_{jx}\}}=0,$ and
$\alpha_{ij}^{11}=4\tr{\{\rho I_{ix}I_{jx}\}}=4\tr{\{\rho
I_{iy}I_{jx}\}}=\alpha_{ij}^{22}$ from the symmetry of the problem.

As a result, the structure of the reduced density matrix of two
spins is given by
\begin{equation}
\label{struc}
\rho_{ij}=\begin{pmatrix} a&0&0&0\\
0&b&x&0\\
0&x&c&0\\
0&0&0&d
\end{pmatrix},
\end{equation}
where
$a=1/4+\frac{\alpha_{ij}^{03}}2+\frac{\alpha_{ij}^{30}}2+\frac{\alpha_{ij}^{33}}4,$
$b=1/4-\frac{\alpha_{ij}^{03}}2+\frac{\alpha_{ij}^{30}}2-\frac{\alpha_{ij}^{33}}4,$
$c=1/4+\frac{\alpha_{ij}^{03}}2-\frac{\alpha_{ij}^{30}}2-\frac{\alpha_{ij}^{33}}4,$
$d=1/4-\frac{\alpha_{ij}^{03}}2-\frac{\alpha_{ij}^{30}}2+\frac{\alpha_{ij}^{33}}4$
and $x=\frac{\alpha_{ij}^{11}}2.$ We calculate the coefficients
$\alpha_{ij}^{03}$, $\alpha_{ij}^{30}$, $\alpha_{ij}^{33}$ and
$\alpha_{ij}^{11}$ below.

\subsection{An open alternating chain with an odd number of spins}

In order to calculate the coefficients $\alpha_{ij}^{03}$,
$\alpha_{ij}^{30}$, $\alpha_{ij}^{33}$ and $\alpha_{ij}^{11}$ we use
the diagonal form of the fermion Hamiltonian of \eq{fham} of the
system. One finds that
\begin{equation}
\label{al00} \alpha_{ij}^{00}=1/4,
\end{equation}
\begin{equation}
\label{al30} \alpha_{ij}^{30}=\tr{\{\rho I_{iz}\}}
=\sum_pu_{ip}^2\frac{e^{-\frac{\beta}2\lambda_p}}{1+e^{-\frac{\beta}2\lambda_p}}-\frac12,
\end{equation}
\begin{equation}
\label{al03} \alpha_{ij}^{03}=\tr{\{\rho
I_{jz}\}}=\sum_pu_{jp}^2\frac{e^{-\frac{\beta}2\lambda_p}}{1+e^{-\frac{\beta}2\lambda_p}}-\frac12,
\end{equation}
\begin{multline}
\label{al33}
\alpha_{ij}^{33}=4\tr{\{\rho I_{iz}I_{jz}\}}\\
=4\tr{\{\rho\l c_i^\dag c_ic_j^\dag c_j-\frac12c_i^\dag
c_i-\frac12c_j^\dag c_j+\frac14\r\}}\\
=4\left[\sum_m\sum_{n\neq
m}u_{im}^2u_{jn}^2\l\frac{e^{-\frac{\beta}2\lambda_m}}{1+e^{-\frac{\beta}2\lambda_m}}\r\l\frac{e^{-\frac{\beta}2\lambda_n}}{1+e^{-\frac{\beta}2\lambda_n}}\r\right.\\
+\left(\sum_n u_{in}u_{jn}\right)\sum_m u_{im}u_{jm}\frac{e^{-\frac{\beta}2\lambda_m}}{1+e^{-\frac{\beta}2\lambda_m}}\\
\left.-\sum_m\sum_{n\neq
m}u_{im}u_{in}u_{jn}u_{jm}\l\frac{e^{-\frac{\beta}2\lambda_m}}{1+e^{-\frac{\beta}2\lambda_m}}\r\l\frac{e^{-\frac{\beta}2\lambda_n}}{1+e^{-\frac{\beta}2\lambda_n}}\r\right]\\
-2\l\sum_ku_{ik}^2\frac{e^{-\frac{\beta}2\lambda_k}}{1+e^{-\frac{\beta}2\lambda_k}}+
\sum_pu_{jp}^2\frac{e^{-\frac{\beta}2\lambda_p}}{1+e^{-\frac{\beta}2\lambda_p}}\r+1.
\end{multline}
Our numerical simulations for the spin chains with 9 spins
demonstrate that entanglement appears only between the
nearest--neighbors even if SSI of all spins are taken into account.
The following coefficients are sufficient for our investigations:
\begin{multline}
\label{al11}
\alpha_{i,i+1}^{11}=\alpha_{i,i+1}^{22}=4\tr{\{\rho I_{ix}I_{i+1,x}\}}\\
=\tr\{\rho(c_i^\dag c_{i+1}^\dag+c_i^\dag c_{i+1}-c_i
c_{i+1}^\dag-c_ic_{i+1})\}\\
=2\sum_nu_{in}u_{i+1,n}\frac{e^{-\frac{\beta}2\lambda_n}}{1+e^{-\frac{\beta}2\lambda_n}}.
\end{multline}
Introducing the notations
\begin{multline}
\label{oboz} C_1=\frac{\omega_2-\omega_1}{D_1},\qquad
C_2=\frac{\omega_2+\omega_1}{D_1},\\
L_k=\begin{cases} \frac1{C_1+\sqrt{C_1^2+\Delta_k}},\quad
k=1,\ldots,\frac{N-1}2\\
\frac1{C_1-\sqrt{C_1^2+\Delta_k}},\quad k=\frac{N+3}2,\ldots,N
\end{cases}
\end{multline}
\begin{equation}
\label{f}
f_k=A_k^2=\begin{cases} \frac4{N+1}\l 1-\frac{\Delta_k}{(C_1+\sqrt{C_1^2+\Delta_k})^2+\Delta_k}\r,\\ k=1,\ldots,\frac{N-1}2\\
\frac4{N+1}\l
1-\frac{\Delta_k}{(C_1-\sqrt{C_1^2+\Delta_k})^2+\Delta_k}\r,\\
k=\frac{N+3}2,\ldots,N
\end{cases}
\end{equation}
\begin{equation}
\label{R}
R_k=A_k^2\l\frac{D_1}{\lambda_k-2w_1}\r^2=\begin{cases} \frac4{N+1}\frac1{(C_1+\sqrt{C_1^2+\Delta_k})^2+\Delta_k},\\
k=1,\ldots,\frac{N-1}2\\
\frac4{N+1}\frac1{(C_1-\sqrt{C_1^2+\Delta_k})^2+\Delta_k},\\
k=\frac{N+3}2,\ldots,N
\end{cases}
\end{equation}
\begin{equation}
\label{e} \epsilon_k=\begin{cases} C_2+\sqrt{C_1^2+\Delta_k},\quad
k=1,\ldots,\frac{N-1}2\\
C_2-\sqrt{C_1^2+\Delta_k},\quad k=\frac{N+3}2,\ldots,N
\end{cases}
\end{equation}
and
\begin{equation}
g(\epsilon_k)=\frac{e^{-\tau\epsilon_k}}{1+e^{-\tau\epsilon_k}},\quad
\tau=\frac{\beta D_1}2,
\end{equation}
we obtain explicit expressions for the coefficient
$\alpha_{i,i+1}^{33}.$ Since the alternating chain is non--symmetric
we obtain the different expressions for the coefficients at even and
odd $i.$ At even $i$ the coefficient $\alpha_{i,i+1}^{33}$ is
\begin{widetext}
\begin{multline}
\label{a33alt}
\alpha_{i,i+1}^{33}=4\left[\sum_{m\neq\frac{N+1}2}\sum_{\substack{n\neq
m\\n\neq\frac{N+1}2}}f_m R_n\right.\sin^2{\l\frac{\pi mi}{N+1}\r}
S_{\delta}^2(n)g(\epsilon_m)g(\epsilon_n)+\sum_{m\neq\frac{N+1}2}f_mB^2(-\delta)^{N-i-1}\sin^2{\l\frac{\pi
mi}{N+1}\r}g\l\frac{2w_1}{D_1}\r g(\epsilon_m)\\
+\left(\sum_n f_nL_nS_{\delta}(n)\sin{\l\frac{\pi ni}{N+1}\r}\right)
\sum_mf_mL_mS_{\delta}(m)\sin{\l\frac{\pi
mi}{N+1}\r}g(\epsilon_m)\\
-\sum_{m\neq\frac{N+1}2}\sum_{\substack{n\neq
m\\n\neq\frac{N+1}2}}f_m
f_nL_nL_mS_{\delta}(m)S_{\delta}(n)\sin{\l\frac{\pi mi}{N+1}\r}
\sin{\l\frac{\pi ni}{N+1}\r}g(\epsilon_m)
\left.g(\epsilon_n)\lefteqn{\phantom{\sum_{\substack{n\neq
m\\n\neq\frac{N+1}2}}}}\right]\\-2
\left\{\sum_{k\neq\frac{N+1}2}f_k\right.\sin^2{\l\frac{\pi
ki}{N+1}\r}g(\epsilon_k)
\left.+\sum_{p\neq\frac{N+1}2}R_pS_{\delta}^2(p)g(\epsilon_p)\lefteqn{\phantom{\sum_{m\neq\frac{N+1}2}}}
+B^2(-\delta)^{N-i-1}g\l\frac{2w_1}{D_1}\r\right\}+1,
\end{multline}
\end{widetext}
where $S_{\delta}(m)=\left[\delta\sin{\l\frac{\pi
mi}{N+1}\r}+\sin{\l\frac{\pi m(i+2)}{N+1}\r}\right].$

At odd $i$ the coefficient $\alpha_{i,i+1}^{33}$ is
\begin{widetext}
\begin{multline}
\label{a33alt}
\alpha_{i,i+1}^{33}=4\left[\sum_{m\neq\frac{N+1}2}\sum_{\substack{n\neq
m\\n\neq\frac{N+1}2}}f_n R_m\right.\sin^2{\l\frac{\pi
n(i+1)}{N+1}\r}
Q_{\delta}^2(m)g(\epsilon_m)g(\epsilon_n)+\sum_{n\neq\frac{N+1}2}f_nB^2(-\delta)^{N-i}\sin^2{\l\frac{\pi
n(i+1)}{N+1}\r}g\l\frac{2w_1}{D_1}\r g(\epsilon_n)\\
+\left(\sum_n f_nL_nQ_{\delta}(n)\sin{\l\frac{\pi
n(i+1)}{N+1}\r}\right) \sum_mf_mL_mQ_{\delta}(m)\sin{\l\frac{\pi
m(i+1)}{N+1}\r}g(\epsilon_m)\\
-\sum_{m\neq\frac{N+1}2}\sum_{\substack{n\neq
m\\n\neq\frac{N+1}2}}f_m
f_nL_nL_mQ_{\delta}(m)Q_{\delta}(n)\sin{\l\frac{\pi m(i+1)}{N+1}\r}
\sin{\l\frac{\pi n(i+1)}{N+1}\r}g(\epsilon_m)
\left.g(\epsilon_n)\lefteqn{\phantom{\sum_{\substack{n\neq
m\\n\neq\frac{N+1}2}}}}\right]\\-2
\left\{\sum_{k\neq\frac{N+1}2}R_kQ_{\delta}^2(k)g(\epsilon_k)\lefteqn{\phantom{\sum_{m\neq\frac{N+1}2}}}
+B^2(-\delta)^{N-i}g\l\frac{2w_1}{D_1}\r+\sum_{p\neq\frac{N+1}2}f_p\sin^2{\l\frac{\pi
p(i+1)}{N+1}\r}g(\epsilon_p)\right\}+1,
\end{multline}
\end{widetext}
where $Q_{\delta}(m)=\left[\delta\sin{\l\frac{\pi
m(i-1)}{N+1}\r}+\sin{\l\frac{\pi m(i+1)}{N+1}\r}\right].$ Explicit
expressions for the coefficients $\alpha_{i,i+1}^{03}$,
$\alpha_{i,i+1}^{30}$ and $\alpha_{i,i+1}^{11}$ can be obtained in a
similar way.

\subsection{The open homogeneous chain}

The expressions of the previous Section simplify for homogeneous
chains and can be written as follows
\begin{equation}
\label{a03h} \alpha_{ij}^{03}=\frac2{N+1}\sum_k\sin^2{\l\frac{j\pi
k}{N+1}\r}g(\epsilon_k)-\frac12
\end{equation}
\begin{equation}
\alpha_{ij}^{30}=\frac2{N+1}\sum_k\sin^2{\l\frac{i\pi
k}{N+1}\r}g(\epsilon_k)-\frac12
\end{equation}
\begin{multline}
\alpha_{ij}^{33}=\frac{16}{(N+1)^2}\left\{\sum_k\sum_{p\neq
k}\sin^2{\l\frac{i\pi k}{N+1}\r}\right.\\sin^2{\l\frac{j\pi
p}{N+1}\r}g(\epsilon_k)g(\epsilon_p)\\
-\sum_k\sum_{p\neq k}\sin{\frac{i\pi k}{N+1}}\sin{\frac{i\pi
p}{N+1}}\\
\left.\sin{\frac{j\pi k}{N+1}}\sin{\frac{j\pi
p}{N+1}}g(\epsilon_k)g(\epsilon_p)\right\}\\
-\frac4{N+1}\left(\sum_k\sin^2{\l\frac{i\pi
k}{N+1}\r}g(\epsilon_k)\right.\\
+\left.\sum_p\sin^2{\l\frac{j\pi p}{N+1}\r}g(\epsilon_p)\right)+1,
\end{multline}
\begin{multline}
\label{a11h} \alpha_{i,i+1}^{11}=\frac4{N+1}\sum_k\sin{\l\frac{i\pi
k}{N+1}\r}\sin{\l\frac{(i+1)\pi k}{N+1}\r}g(\epsilon_k),
\end{multline}
where $D_1=D_2=D,$ and the one-fermion spectrum $\epsilon_k$ is
defined as
$$
\epsilon_k=2\cos{\frac{\pi k}{N+1}}+\frac{2\omega_0}D,\quad (k=1,
2,\ldots, N).
$$
We emphasize again that \eqs{a03h}{a11h} are valid for all~N.

\section{measure of entanglement}

In this paper we restrict our attention to the analysis of the
entanglement between two arbitrary spins in the chain. Here we
consider the concurrence as a measure of entanglement~\cite{woot}.

Let $A$ and $B$ be a pair of qubits, and let the density matrix of
the pair be $\rho_{AB}$ which may be pure or mixed. Then the
"spin-flipped" density matrix is determined as
\begin{equation}
\label{flip}
\tilde{\rho}_{AB}=(\sigma_y\otimes\sigma_y)\rho_{AB}^*(\sigma_y\otimes\sigma_y)
\end{equation}
where the asterisk denotes complex conjugation in the standard basis
$\{|00\rangle,|01\rangle,|10\rangle,|11\rangle\}$ and the Pauli
matrix $\sigma_y=2I_y.$ The concurrence of the two--spin system with
 the density matrix $\rho_{AB}$ is~\cite{woot}
\begin{multline}
C_{AB}=max\{0, 2\lambda-\lambda_1-\lambda_2-\lambda_3-\lambda_4\}\\
\lambda=max\{\lambda_1,\lambda_2, \lambda_3, \lambda_4\}
\end{multline}
where $\lambda_1,$ $\lambda_2,$ $\lambda_3,$ and $\lambda_4$ are the
square roots of the eigenvalues of the product
$\rho_{AB}\tilde{\rho}_{AB}.$ Since both $\rho_{AB}$ and
$\tilde{\rho}_{AB}$ are positive operators, it follows that their
product, though non--Hermitian, has only real and non--negative
eigenvalues~\cite{horn}.

Using \eq{struc} one can find the expressions of $\lambda_1,$
$\lambda_2,$ $\lambda_3,$ $\lambda_4$ as follows
\begin{multline}
\label{lam}
\lambda_1=\lambda_4=\sqrt{ad},\\
\lambda_{2,3}=|x\pm\sqrt{bc}|.
\end{multline}
Expressions for \eq{lam} yield all information necessary for the
analysis of the pairwise entanglement for such chains.

\section{numerical analysis of the concurrence in spin pairs}

As mentioned above, our numerical calculations in the nine--spin
chain with the Hamiltonian~of~\eq{iham} lead to the conclusion that
the concurrence is non--zero for nearest neighbors only. We focus
below on entanglement of the nearest neighbors in the open
alternating chain with the XY-Hamiltonian in the case of zero Larmor
frequencies.
\begin{figure}[h]
\begin{center}
\includegraphics[width=8.6cm]{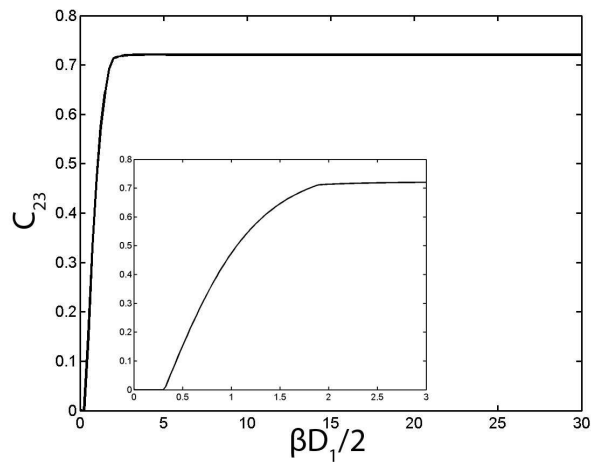}
\end{center}
\caption{The concurrence of spins $2$ and $3$ versus the temperature
for the number of spins $N=101$ and $D_2/D_1=1.5.$ The inset
displays the concurrence versus the temperature on a greater scale.}
\label{2-3alt}
\end{figure}

Using \eqs{al00}{a11h},~(\ref{lam}) we investigate numerically the
dependence of the concurrence, $C_{12},$ of the first and second
spins on the temperature. Figure~\ref{2-3alt} shows that
entanglement appears at $\beta D_1\approx1.$ This corresponds to the
temperature $T\approx0.5 \mu K$ at $D_1\approx2\pi\cdot10^4 c^{-1}.$
The temperature at which entanglement emerges in the pairs of
neighboring spins depends on the ratio of the SSI constants, chain's
length and the distance from ends of the chain. It is interesting to
note that ordered states of nuclear spins were observed at
microkelvin temperatures~\cite{abraham}.

\begin{figure}[t]
\begin{center}
\includegraphics[width=8.6cm]{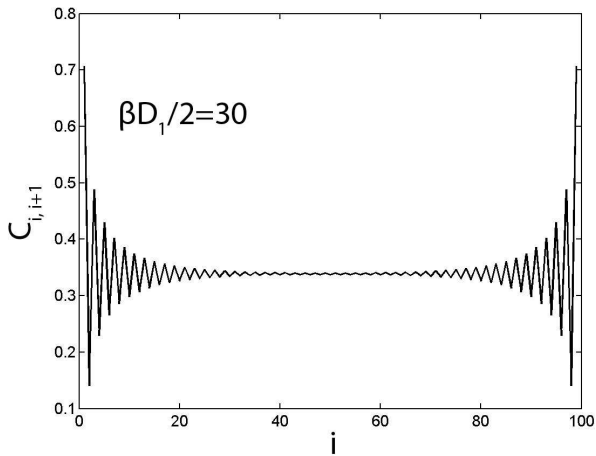}
\end{center}
\caption{The nearest-neighbor concurrence versus the site number in
the homogeneous chain for $N = 100.$ The concurrence oscillates with
a two--site period. The oscillations decay when the spin pair is far
from the ends of the homogeneous chain.} \label{h-osc}
\end{figure}

\begin{figure}
\begin{center}
\includegraphics[width=8.6cm]{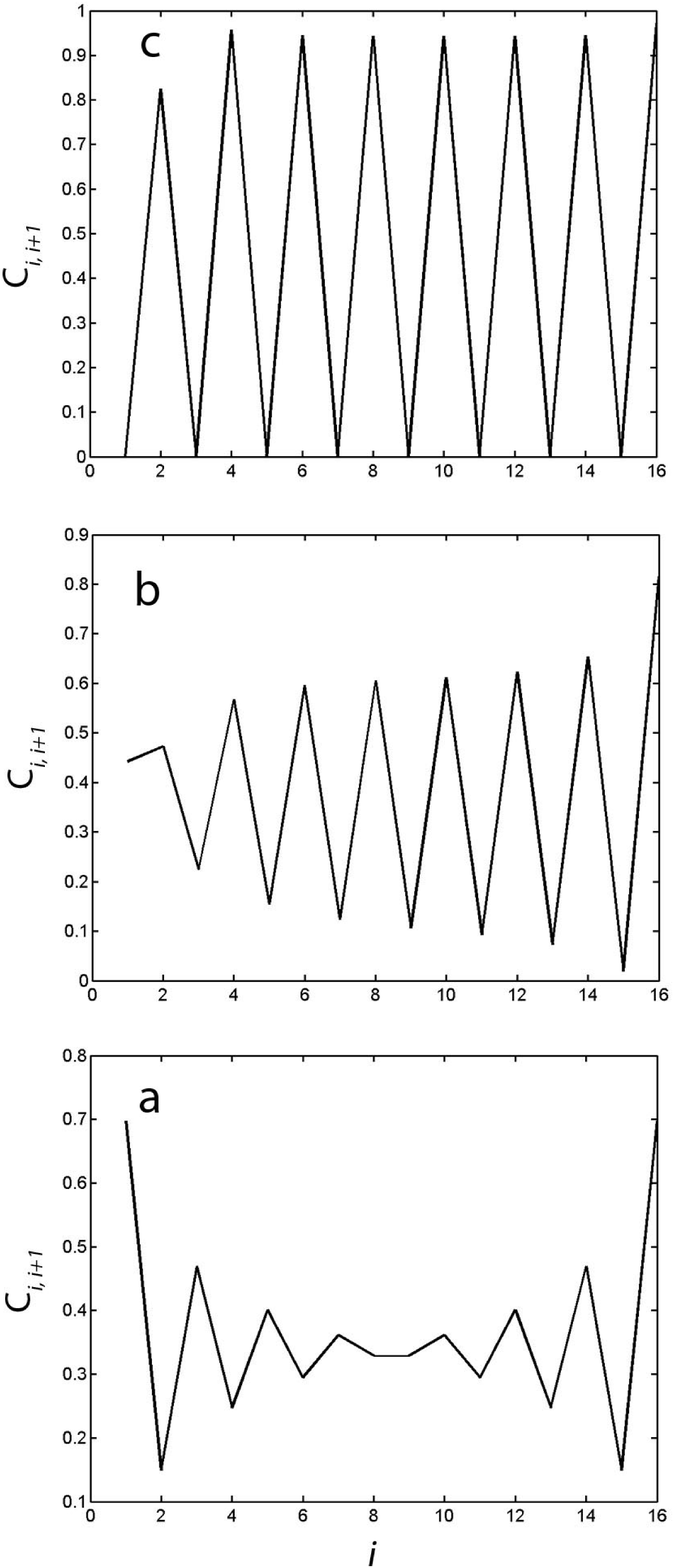}
\end{center}
\caption{The nearest-neighbor concurrence versus the site number for
$N = 17$ and $\beta D_1/2=30$ at the difference ratios of the SSI
coupling constants; a\,:\,$D_2/D_1=1;$ b\,:\,$D_2/D_1=1.17;$
c\,:\,$D_2/D_1=3.$} \label{a-osc}
\end{figure}

\begin{figure}[h]
\includegraphics[width=8.6cm]{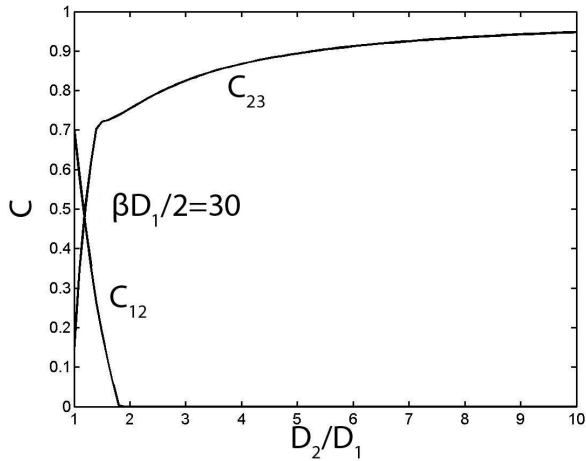}
\caption{The nearest-neighbor concurrence of the pair of spins 1 and
2 and of the pair of spins 2 and 3 versus the ratios of the SSI for
$N = 55$ and for $\beta D_1/2=30$.} \label{ratio}
\end{figure}

Numerical results show (Fig.~\ref{h-osc}) that entanglement
oscillates with the two--site period. The qualitative explanation of
the numerical results is the following.  Qubits 1 and N are situated
at the chain ends. Since  entanglement is non--zero only between the
nearest--neighbors, the pair of qubits 1 and 2 and the pair of
qubits N-1 and N have the maximal pairwise entanglement for the
homogeneous chain. Spin 2 can be entangled both with spin 1 and with
spin 3. Since spin 2 is strongly entangled with spin 1, the
entanglement of spins 2 and 3 is weaker. As a result, spin 3 is
strongly entangled with spin 4, etc. This explains the oscillator
behavior of the concurrence displayed in Fig.~\ref{h-osc}.
The oscillations decay when the spin pair is far from the ends of
the homogeneous chain. These oscillations do not decay for
alternating open chains as it is displayed in Fig.~\ref{a-osc} at
different ratios of the coupling constants, $D_2/D_1.$
The oscillation of the entanglement from zero to values close to one
is due to different coupling constants between spins in the
alternating chain. In fact we have here the dimerised spin chain
which can be considered qualitatively as a set of non--interacting
spin pairs  at $D_2/D_1\geq2.$ The dependence of the concurrence on
the ratio of the SSI is shown in Fig.~\ref{ratio}. It is worth to
notice that it is impossible to observe the influence of ends of the
chain  on the entanglement already at $D_2/D_1=1.5$.

\begin{figure}
\includegraphics[width=8.6cm]{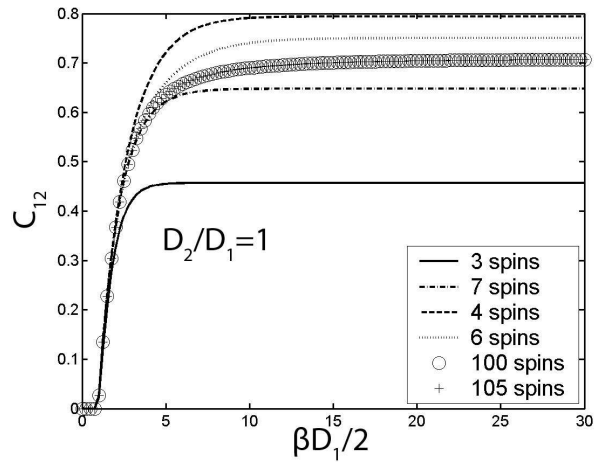}
\caption{The concurrence of spins 1 and 2 versus the temperature for
the number of spins $N=3, 4, 6, 7, 100, 105.$ The behavior of the
concurrence $C_{12}$ is different for chains with even and odd N.}
\label{1-2lt}
\end{figure}

\begin{figure}[t]
\includegraphics[width=8.6cm]{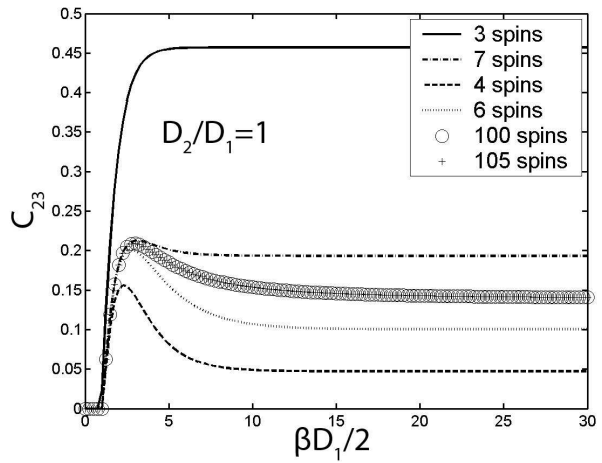}
\caption{The concurrence of spins 2 and 3 versus the temperature for
the number of spins $N=3, 4, 6, 7, 100, 105.$} \label{2-3lt}
\end{figure}

In Figs.~\ref{1-2lt}~--~\ref{2-3lt} the dependence of the
concurrences $C_{12}$ and $C_{23}$ on the length of the homogeneous
chain consisting of an odd (even) number of spins is displayed. The
concurrence $C_{12}$ decreases at even N and increases at odd N when
N increases. For the concurrence $C_{23}$ the situation is opposite.
In order to explain these results it is necessary to take into
account the influence of the ends of the chain on the entangled
states.
As a result, we obtain that the ends of the chain yield opposite
contributions to the entanglement for the chain of an odd number of
spins. For example, one can easily find in a chain consisting of 5
spins that spin 5 leads to a decrease of the entanglement of spins 1
and 2. This effect is diminished when the odd number of spins
increases and then the entanglement of spins 1 and 2 increases. On
the contrary, the second end of the chain increases the entanglement
of spins 1 and 2 at an even number of spins. This leads to the
calculated decrease of the concurrence (see Fig.~\ref{1-2lt}) when
the even number of spins increases. Analogously, it is possible to
explain the growth (fall) of the concurrence of the entanglement of
spins 2 and 3 in the dependence on the length of the chain
consisting of an even (odd) number of spins~(see Fig.~\ref{2-3lt}).
The maximums of the concurrence in Fig.~\ref{2-3lt} have a simple
qualitative explanation. The growth of the concurrence up to its
maximal value is determined by the decrease of the partial
suppression of the SSI due to the flip--flop transitions at low
temperatures. The state of spins 2 and 3 approaches to a separable
one at low temperatures when the spins are trying to orient parallel
to the local fields. Entanglement of such a state decreases when the
temperature decreases.

\section{conclusion}
In the present work we investigate entanglement in alternating open
chains. The developed methods of the diagonalization of the
XY-Hamiltonian of open alternating spin chains allow us to study the
pairwise entanglement for different parameters of the chain and its
temperature.

Similar methods can be applied for an investigation of the pairwise
entanglement in open chains with periodic coupling
constants~\cite{kfeld}. Methods of exact diagonalization of the
XY-Hamiltonian in such chains~\cite{kfeld} allow a solution of
different problems of quantum information theory for  models of
quantum registers taking into account the qubit addressing. However,
the qubit addressing requires a large difference of the Larmor
frequencies of different spins in comparison with their SSI coupling
constants. Our preliminary calculations demonstrate that the
concurrence is close to zero at such conditions and the entangled
states do not appear. The question is a subject of our further
investigations.

\section{acknowledgments}
We thank Professor D. E. Feldman, Professor A. K. Khitrin and Dr.
J.-S. Lee for stimulating discussions.

\end{document}